# Digital analysis of early color photographs taken using regular color screen processes

Jan Hubička[1], Linda Kimrová[2], Kenzie Klaeser[3], Sara Manco[4], Doug Peterson[3]

[1]Department of Applied Mathematics, Charles University; Šechtl and Voseček Museum of Photography; SUSE LINUX s.r.o.

[2]Charles University

[3]Digital Transitions

[4]National Geographic Society,

Contact: Jan Hubička, honza.hubicka@gmail.com

**Abstract**

Some early color photographic processes based on special color screen filters pose specific challenges in their digitization and digital presentation. Those challenges include dynamic range, resolution, and the difficulty of stitching geometrically–repeating patterns. We describe a novel method used to digitize the collection of early color photographs at the National Geographic Society which makes use of a custom open-source software tool to analyze and precisely stitch regular color screen processes.

**Keywords:** Early color photography, National Geographic Society, Dufaycolor, Finlay Colour plate, Paget Colour plate, digitization of color photographs

**Introduction**

Several commercially produced processes of early color photography (used in 1890s to 1950s) were based on special *color screen* filters consisting of small *patches* of red, green, and blue color (Fig 3a). The filter was either an integral part of the photographic film or attached on top of the panchromatic negative before taking the picture. After development, monochromatic transparency was produced. When registered with the color screen filter it yields a color photograph. Most common examples of these processes include Autochrome, Agfacolor, Dufaycolor, Paget and Finlay color plates. The first two use random (or stochastic) color patterns, while the remaining use periodic geometrical patterns similar to modern Bayer filter used in digital cameras (Lavédrine and Gandolfo, 2013), (Pénichon, 2013). We call those processes *regular color screen processes*.

Digitization of color screen processes has high demands on the quality of digital captures. It is necessary to digitize large format originals in multiple captures and later stitch individual tiles to a high-resolution image. This task has turned out to be impossible to do with usual panorama stitching tools since small errors in the registration of individual tiles lead to easily noticeable artifacts visible in the regular color screen pattern.

We describe practical experience with digitizing collections of photographs taken using Dufaycolor film, Paget, and Finlay color plates at the National Geographic Society by digitization service provider Digital Transitions and using specialized open-source tool Color-Screen.

**Early color photography collection at National Geographic Society**

Founded in 1888, the National Geographic Society is a 501 c3 non-profit that uses the power of science, exploration, education and storytelling to illuminate and protect the wonder of our world. Photography is a major part of that history, including contributing to technical photographic advancements in addition to its memorable images seen by the public.

The Society was no stranger to color photography when the Autochrome first arrived. Invented by Lumière brothers, the Autochrome was the first industrialized color photographic process – before this nearly all color photographs were made so by hand tinting. The National Geographic Magazine had already been publishing those hand tinted photos in color, and editors were incredibly excited for a photograph that captured colors seen by human eyes. Though the Society was first interested in Autochromes, it quickly acquired many examples of other processes Agfacolor, Dufaycolor, Finlay and Paget color plates.

In total, the Society has one of the largest collections of screen plate photography in the world at over 15,000 plates documenting many parts of the globe. These photographs were used by the Geographic to illustrate magazine stories until the invention of Kodachrome film in the mid 1930s.

In 2020, the Society began the Early Color Photography Conservation and Digitization Project made possible in part by a major grant from the National Endowment for the Humanities: *Democracy demands wisdom*. The project was broken down into two phases: *Phase One – Conservation* and *Phase Two – Digitization*.

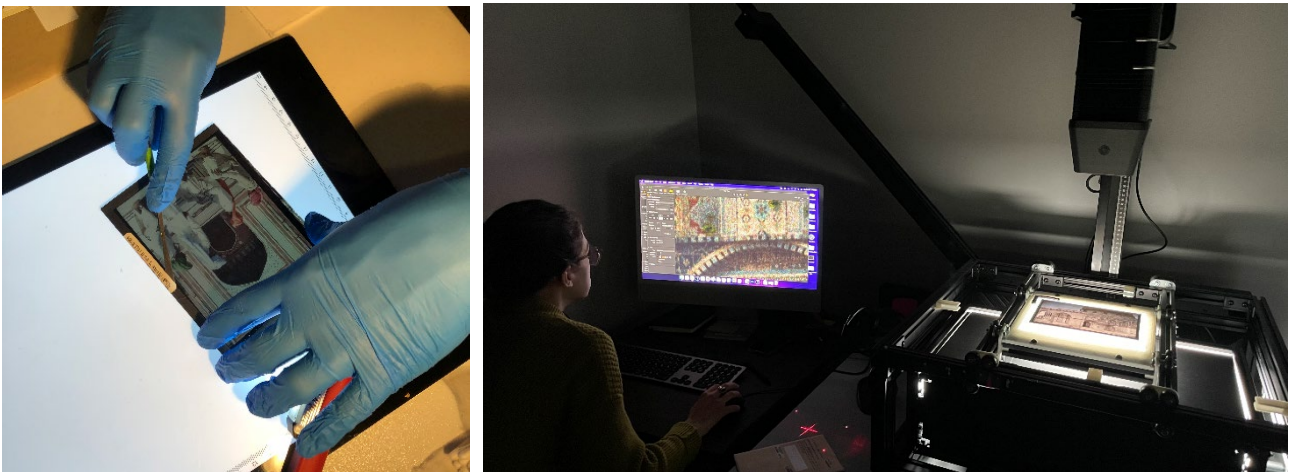

Fig. 1 – Conservation (left) and digitization (right)

During Phase One, the Society collaborated with the Conservation Center for Art and Historic Artifacts (CCAHA) to conserve each photograph in this collection. The objects were cleaned of fingerprints, dust, and pollutants such as cigarette smoke prior to digitization. Just under 400 items were transported to CCAHA for in depth treatment including glass replacement or puzzling back together items broken into several pieces. A particularly important factor in this work was the type of glass used for repair. Borosilicate and plexi were both chosen because they will not create newton rings in the digitized image. As we were conserving the Dufaycolor plates, we discovered they had developed vinegar syndrome. The Dufays were stored near negatives with vinegar syndrome for many years, and the acetic acid seeped through the tape and in between the two pieces of glass in all Dufays to cause the classic vinegar syndrome symptoms: cracking of the film base, warping of the emulsion, a strong vinegar smell, and dyes forming bright purple spots (Siegel, 2003). There is no conservation treatment to reverse this type of deterioration, and the cracking and warping posed some challenges during the digitization.

During Phase Two, the Society collaborated with Digital Transitions Heritage (DT) to digitize each photograph. In order to create the most detailed image, each object was digitized in sections and stitched together in the post-process workflow. Every stitched file was quality checked with a 3–5% targeted fail rate ensuring accurate color reproduction and high quality stitching in the final product. Failed stitching exhibited at least one of the following qualities: a stitch line throughout the photograph, color and/or tonal differences between the tiles, and misaligned content of the image. For those that failed, manual stitching fixed most of the errors and images containing large blocks of color (such as a blue cloudless sky) were some of the hardest to stitch. But as we discuss below the Dufaycolor photographs presented their own unique challenge in the stitching process.

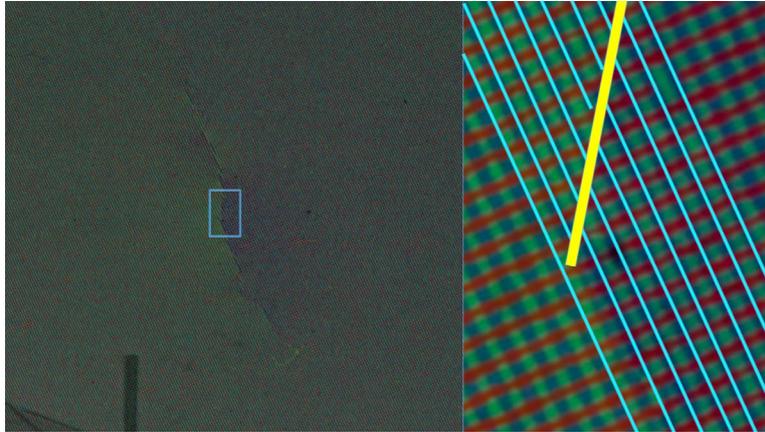

Fig. 2 – Stitching errors are visible as both tears in the geometrical grid as well as color changes.

The Society will store in its DAMS (Digital Asset Management System, which is backed up in the cloud) each tile, the stitched master TIFF, the content reproduction JPEG, and delivery JPEG for each photographic object. Though the content reproduction and the delivery jpegs will be accessed by users the most, each tile is kept as a master reference and for possible restitching in the future without having to re-digitize the original photograph. Each file is embedded with IPTC metadata notes about the stitching and other post-processing information unique to this project.

**Digitization Technical Parameters**

Digitization of these materials requires very high image quality, especially if the results will be used in computational imaging techniques such as stitching or color screen demosaicing. While an exhaustive list of image quality parameters and digitization technique is out of scope for this paper, three such parameters are of special interest: resolution, dynamic range, and geometry.

It is vital that the resolution of the capture out resolve the physical pattern. Otherwise, the screen pattern cannot be cleanly resolved by the eye, and may be misdetected by an algorithm. Digital Transitions conducted research with National Geographic in 2019 which established baseline PPI requirements for each early color process in their collection. By far the most demanding was Dufaycolor which required 4400 ppi to cleanly resolve the screen. This means that for a 5 × 7 inch plate (with 0.25 inch padding) roughly 22,000 × 33,000 pixels is required for the plate, which means even a 150 megapixel capture device must capture the plate in tiles. It is also important that these pixels represent sharp and homogenous detail across the frame; that is, in signal processing terms, a high sampling efficiency is required throughout the frame. We used the ISO 19264 and FADGI 2023 imaging guidelines to select appropriate SFR10, SFR50, and SFRmax aim values (Rieger *et al.*, 2023).

It is also important that the captures be high in dynamic range (high bit depth + low noise). This is primarily because such plates are often very dense (high optical density; low transmittance; dark) and because it is standard archival practice to set exposure such that the white point is the naked light source. However, it is also because the colorants of the screen pattern closely align to the Bayer pattern of the sensor, and therefore each plate pixel is primarily recorded only with its corresponding digital sensor pixel. For example, a very dark red line of a Dufay plate will only be properly seen by the red pixels of the digitization system's Bayer filter; the green and blue pixels of the digitization system will record no meaningful signal (only the noise floor). This is always a problem in digital imaging, but the highly saturated primary-color nature of the mosaic plates is a worst-case scenario.

Finally, it is important to minimize systematic geometric distortions. The Color-Screen tool will account for all manner of distortions including non-planarity and lens distortion, and has even been adapted to account for scanner-drag (uneven spatial stepping due to variation in the movement of a scanner that moves a scan head during digitization). However, the more distortions in the imaging system the wider the tolerances Color-Screen must allow, which increases the chance of failures.

Based on these technical requirements DT selected the DT Atom with a 150mp DT RCam, and built a custom high-precision XY accessory. We calibrated the system with a DT LaserAlign and checked the homogeneity of detail using an ISA Golden Thread Film Target. This system features a lockable focus ring, which eliminates focus drift (a common occurrence with general purpose cameras when you point the lens downward). A DT technician, Kenzie Klaeser, familiar with the system, operated the system throughout the project until its conclusion in September 2023, assisted by various members of the National Geographic Society.

**Color-Screen tool**

Color-Screen[1] is an open-source tool for digital processing of scans of photographs taken using color screen processes. In 2010 Mark Jacobs and Jan Hubička noticed scans of Finlay color negatives in the American Colony Collection at the Library of Congress. After many unsuccessful attempts to register the scans with viewing screen in GIMP and Photoshop, Jan implemented a prototype of a specialized tool. First color reconstructions were shown at 2011 exhibition "The American Colony in Jerusalem and the Matson Photo Service" at the Šechtl and Voseček Museum of Photography in Tábor, Czech Republic. Due to lack of scans of similar negatives, the project was put on hold until 2022 when Linda Kimrová joined the project and new collections of similar early color photographs were found, see (Barker *et al.*, 2022) for list.

Accurate color reconstruction from monochromatic negatives requires understanding of spectra of original dyes. These are known for Dufaycolor (Cornwell-Clyne, 1951). For Autochrome modern analysis was performed based on recreated dyes (Lavédrine, 1992), (Casella and Tsukada, 2012). We are not aware of similar study of Finlay and Paget screens. Modern spectral measurements (Trumpy *et al.*, 2021) provides useful data and can be used to simulate aging of the image as well as making "educated guess" on the original colors.

Color-Screen is useful also when digitizing final transparencies with color screens attached. Since dyes used in the screen filters are not very stable, most positives appear very faded. In some cases, the registration between the black and white transparency and the screen has been lost which leads to complete or partial color loss of the image. Unlike for later subtractive color films, in the color screen processes the original image is recorded in silver-gelatine transparency which is stable and can be digitized in infrared spectra (dyes used in the viewing screen are transparent in infrared scans). This enables Color-Screen to accurately reconstruct original colors from scans.

Even with a specialized GUI, registering screens manually is a time-consuming process (main challenge poses numerous geometric errors in the scans as well as aging of the film base or flaking of the emulsion). Work on automatic analysis of screens based on the scans started in 2022.

---

[1] https://github.com/janhubicka/Color-Screen/wiki

**Extending Color-Screen to stitch high resolution scans**

While not originally intended to be used to stitch scans, it was clear that algorithms used to analyze geometry of the color screens can be useful to assist image stitching software.

Our tool works in three basic steps: 1) it analyzes every tile independently, 2) it solves the stitching problem and 3) renders individual tiles of the output image to be blended by separate software.

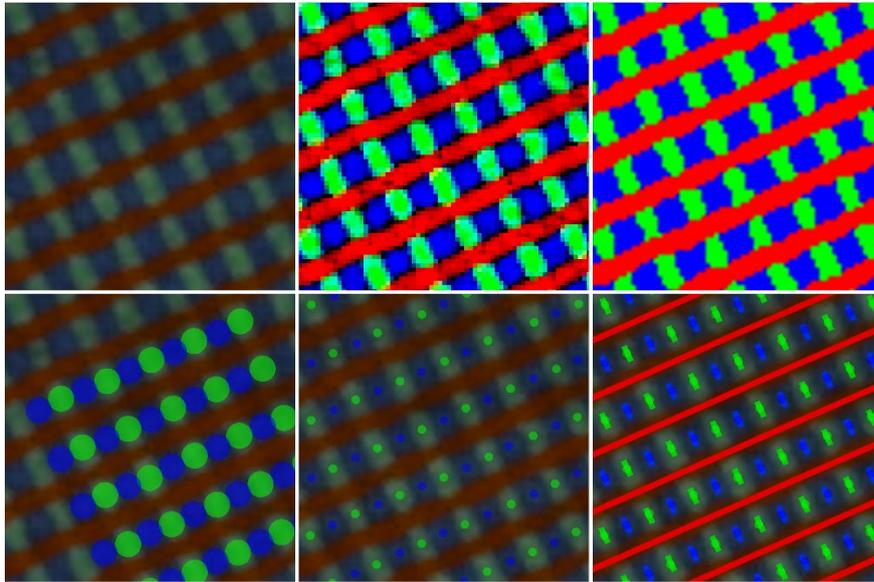

Fig. 3 – Dufaycolor color screen and its analysis. a) original screen, b) screen after analysis of color filters, c) screen after identification of red, green and blue patches, d) detected grid of blue and green patches, e) result of flood fill, f) result of RANSAC mesh construction

The first part of the analysis is most involved and divided into the following steps:

***Color linearization***: Image is converted from input profile to linear RGB. It is important to work with scans close to their raw form without any digital enhancement algorithms applied (such as tone curve, sharpening, noise reduction, color space conversions) except for demosaicing (Fig. 3a).

***Analysis of color filters***: A sample of image in the center of scan is analyzed identifying 10% of most red, blue and green pixels. These are used to build "matrix profile" of the original screen. This is done by estimating 3-dimensional vectors $\vec{d}, \vec{r}, \vec{g}, \vec{b}$. Vector $\vec{d}$ determines black point of the scan, while vectors $\vec{r}, \vec{g}, \vec{b}$ determines color of the red, green and blue filters of the color screen as interpreted by the digital scan. Ideally all samples of red color should be in the form $\vec{d} + \vec{r}x$, where $x$ is a scalar variable. Similarly for blue and green samples. Result is a $4 \times 4$ matrix $M$. Based on the matrix $M$ the image is converted from its original linearized RGB to RGB values which correspond to the estimated profile (Fig. 3b) and unsharp-mask is applied.

***Identification of red, green and blue patches***: Given pixel with color $(r, g, b)$ if $r > |g + b|t$ for a fixed threshold $t$ we consider given pixel to be red. Similarly for green and blue (Fig. 3c).

***Detection of the geometry of color screen***: This step depends on a particular color screen. For Dufaycolor screen first a continuous area of green pixels is found. Because of the organization of the Dufaycolor screen it is expected that neighboring areas (patch) will be blue. Based on the center of mass of the two neighboring patches it estimates the orientation and scale of the screen. Next algorithm looks for a regular grid of $5 \times 5$ of pairs of green and blue patches gradually improving estimates of the screen geometry (Fig. 3d). Analysis Paget and Finlay screens is similar.

*Flood fill*: The parameters determined are not precise enough to register full scan with the color screen, however they are quite good locally. To determine as many color patches as possible the algorithm extending classical flood-fill is used. Based on knowledge of one-color patch and estimated geometry, neighboring patches are detected and recursively also neighbors of the neighbors etc. While not all patches are rendered well on the scans, in a good scan over 95% of patches can be identified. This step is most important for getting good match between the screen and the scan. While it seems easy, it is important to not make off-by-one errors while walking millions of patches and this required significant amount of fine-tuning.

*Mesh construction*: Result of the flood-fill algorithm never fully perfect. Some patches are missing, and the centers of patches are not very well computed. This problem can however be solved by simple statistical tools. We implement standard Random sample consensus (RANSAC) algorithm to determine homography between the color screen and identified color patches. With additional lens correction this produces good results for scans of photographs on glass (Paget and Finlay).

Dufaycolor has film base which is deformed due to its deterioration. To solve this problem, we run RANSAC solver in a grid of $100 \times 100$ points over the area of the scan each time setting weight of individual detected patches to decrease by the distance from the center of the point analyzed. Smoothly interpolating the solution leads to an adaptive model that is flexible enough to follow the deformed grid on film bases and still understand the fact that locally the transformation should correspond to a homographic map.

*Data collection*: Once the location of each color patch is identified, its intensity is determined from the original scan. The result is a mosaiced digital image in the natural resolution of the color screen, each patch corresponding to one pixel, which is later used in the stitching process.

Each of the steps may fail if the algorithm gets confused by some defects or irregularity of the image (such as broken part of the image, hand retouching, very dark area etc.). If Flood fill algorithm fails to analyze important part of the scan the process is repeated starting from different location than the center of the scan.

Once all tiles of the scan are analyzed, the stitching is performed using standard open-source tool cpfind[2] based on a demosaiced version of image produced in data collection (just as in digital photograph). Once matching points between scans are identified Color-Screen applies an inverse mesh transform to identify matching points in the original scan. These can be inserted to an open-source panorama stitching tool Hugin[3].

Pairs of matching points identified are significantly more precise than ones done by general-purpose panorama stitching tool. Yet it turned out that using Hugin to produce final scan is leading to stitching artifacts. In our test samples the average point distance was in the range 0.5–1.3 pixels and maximal distance 5–11 pixels. While this precision would be acceptable for production of normal panorama, with average size of a color patch in the scan being just 5 pixels, such displacements are easily visible and moreover they confuse color blending algorithms (Fig. 2).

Detailed analysis showed that this problem is related to the film flatness. Dufaycolor transparencies were sandwiched between two glass plates. Even in this setup the film flatness is not perfect and since each tile scans the film from a different angle, this leads to complex geometrical errors. Other common source of a problem are vibrations of camera during the capture. For this reason, stitching using homographic transformations and lens corrections fails.

---

[2] https://wiki.panotools.org/Cpfind
[3] https://hugin.sourceforge.io/

Instead of relying on the standard image stitching algorithms we implemented an algorithm that produces geometrically corrected image based on the transformation mesh constructed in the initial stage. This corrects both the deformations caused by digital capture process as well as the deformations caused by aging of the film base.

The final problem to solve was blending of individual tiles. Even with calibrated backlighting, the tiles do not match in color and density. This is caused by the fact that there is a space between the photographic emulsion and color screen and thus different capture angles lead to different colors. We adapted open-source tool multiblend[4] for this purpose.

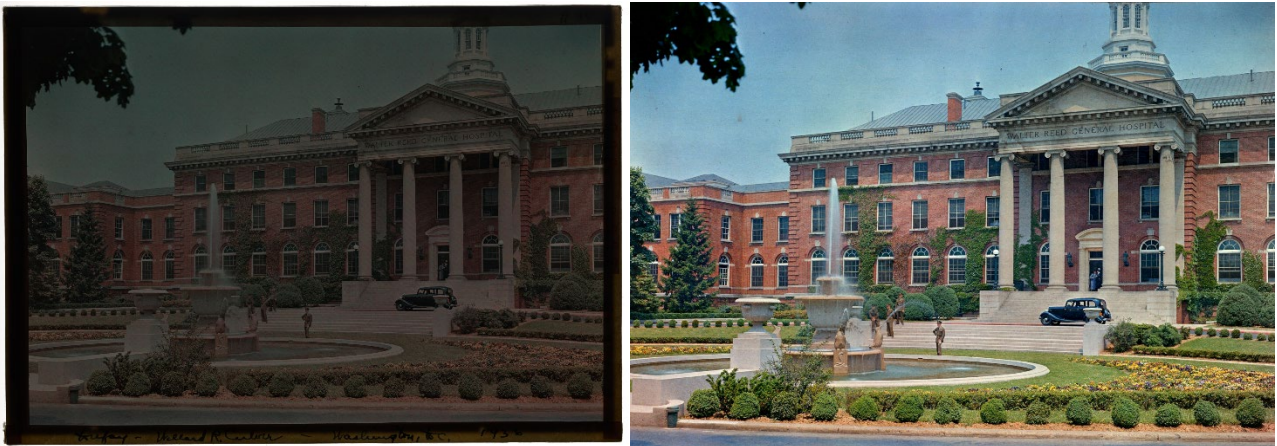

Fig. 4 – Walter Reed General Hospital by Willard Culver, 1936. Successful stitch (left). With demosaic and color reconstruction (right).

**Production Integration**

Color-Screen is a command line tool that handles one photographic image at a time. The large National Geographic Collection required a more scalable implementation. So, Color-Screen was integrated into an imaging pipeline called DT PixelFlow used by Digital Transitions to automate stitching as one of several steps in the production process. Specifically, DT PixelFlow converted the camera raw files to TIFFs, renamed them to comport to Color-Screen's expectations, called Color-Screen, added metadata to the resulting files, and created additional derivatives such as a JPEG that was visually reviewed by a quality control technician, and a TIFF with an automatic brightening adjustment applied to make dense plates more legible.

**Conclusions and Future work**

The primary goal has been to develop a method to digitally capture regular color screen processes in a resolution which faithfully record their mosaiced nature. This is pushing current digital imaging to limits, however the fact that Color-Screen tool can analyze the scan and identify individual color patches serves also as a proof that this goal has been established.

Tools developed have multiple additional uses.

---

[4] https://horman.net/multiblend/

1) Because it precisely analyzes the viewing screen it is possible to apply demosaicing and produce high resolution images that are more fit for printing and viewing on common digital screens than their mosaiced originals.
2) Using the knowledge of spectra of dyes used in Dufaycolor screen it is also possible to attempt careful color reconstructions of the captured images. Ideally this should be based on scans which include red, green, blue and infrared channels. However even without infrared scan a limited-quality reconstruction is possible (Fig. 4)
3) It is possible to identify different variants of color screens used. In the National Geographic Society collection two types of Dufaycolor images are used which differ by size and angle of the individual lines forming the color screen. This difference appears to not be discussed in literature, yet.